\documentclass[aps,twocolumn,amssymb,amsfonts,amsmath,showpacs,final,prl,superscriptaddress]{revtex4-1}

\usepackage{float}
\usepackage[dvips,final]{graphicx}
\usepackage{color}
\usepackage{amsmath}

\newcommand{\ket}[1]{\left|#1\right>}
\newcommand{\bra}[1]{\left<#1\right|}

\newcommand{\nn}{\nonumber\\}

\newcommand{\bea}{\begin{eqnarray}}
\newcommand{\ea}{\end{eqnarray}}
\newcommand{\eea}{\end{eqnarray}}

\newcommand{\sumint}[1]

\begin{document}

\title{Doublon bottleneck in the ultrafast relaxation dynamics of hot electrons in 1\emph{T}-TaS$_{2}$}

\author{I. Avigo}
\affiliation{Fakult\"{a}t f\"{u}r Physik, Universit\"{a}t Duisburg-Essen, Lotharstr.~1, 47057 Duisburg, Germany}

\author{F. Queisser}
\affiliation{Fakult\"{a}t f\"{u}r Physik, Universit\"{a}t Duisburg-Essen, Lotharstr.~1, 47057 Duisburg, Germany}
\affiliation{Helmholtz-Zentrum Dresden-Rossendorf, Bautzner Landstr.~400, 01328 Dresden, Germany}
\affiliation{Institut f{\"u}r Theoretische Physik, Technische Universit\"at Dresden, 01062 Dresden, Germany}

\author{P. Zhou}
\affiliation{Fakult\"{a}t f\"{u}r Physik, Universit\"{a}t Duisburg-Essen, Lotharstr.~1, 47057 Duisburg, Germany}

\author{M. Ligges}
\altaffiliation{Current address: Fraunhofer IMS, 47057 Duisburg, Germany}
\affiliation{Fakult\"{a}t f\"{u}r Physik, Universit\"{a}t Duisburg-Essen, Lotharstr.~1, 47057 Duisburg, Germany}

\author{K. Rossnagel}
\affiliation{Institut f\"{u}r Experimentelle und Angewandte Physik, Christian-Albrechts-Universit\"{a}t zu Kiel, 24098 Kiel, Germany}
\affiliation{Ruprecht-Haensel-Labor, Christian-Albrechts-Universit\"{a}t zu Kiel und Deutsches Elektronen-Synchrotron DESY, 24098 Kiel and 22607 Hamburg, Germany}
\affiliation{Deutsches Elektronen-Synchrotron DESY, 22607 Hamburg, Germany}

\author{R. Sch\"{u}tzhold}
\email[]{r.schuetzhold@hzdr.de}
\affiliation{Fakult\"{a}t f\"{u}r Physik, Universit\"{a}t Duisburg-Essen, Lotharstr.~1, 47057 Duisburg, Germany}
\affiliation{Helmholtz-Zentrum Dresden-Rossendorf, Bautzner Landstr.~400, 01328 Dresden, Germany}
\affiliation{Institut f{\"u}r Theoretische Physik, Technische Universit\"at Dresden, 01062 Dresden, Germany}

\author{U. Bovensiepen}
\email[]{uwe.bovensiepen@uni-due.de}
\affiliation{Fakult\"{a}t f\"{u}r Physik, Universit\"{a}t Duisburg-Essen, Lotharstr.~1, 47057 Duisburg, Germany}

\date{\today}

\begin{abstract}
Employing time-resolved photoelectron spectroscopy we analyze the relaxation dynamics of hot electrons in the charge density wave / Mott material 1\emph{T}-TaS$_2$.
At 1.2~eV above the Fermi level we observe a hot electron lifetime of  $12\pm5$~fs in the metallic state and of $60\pm10$~fs in the broken symmetry ground state -- a direct consequence of the reduced phase space for electron-electron scattering determined by the Mott gap. Boltzmann equation calculations which account for the interaction of hot electrons in a Bloch band with a doublon-holon excitation in the Mott state provide insight into the unoccupied electronic structure in the correlated state.
% and confirm the $U/J\sim1$ limit.
\end{abstract}

\maketitle

The lifetime of an excited, hot electron is determined by the imaginary part of the self energy and for bulk metals \cite{chulkov_2006,Bauer15} and semiconductors \cite{Shah99} a comprehensive understanding has been developed. For materials with strong electron correlations such insight is missing because (i) the electronic structure is considerably more complex and (ii) the excitation of interest may modify the electronic structure hosting it. In recent years established experimental and theoretical approaches, which investigate the thermal equilibrium, were complemented by methods which access non-equilibrium states of matter in the time domain. While part of the activity aims at states and properties which exist out of equilibrium \cite{ichikawa2011,Foerst2011,stojchevska14,cui2014,Basov2017}, also new tools for the analysis of excitations in strongly correlated materials were introduced \cite{Bovensiepen12,gianetti2016,gerber17,aoki14,parham17}. For low energy excitations up to 100~meV differences between the single-particle and the population lifetimes occur due to carrier relaxation and multiplication \cite{yang15}. Such differences are absent for higher energies \cite{boger_2002}. Due to the complexity of the problem rather simple models like the Falicov-Kimball model \cite{eckstein_2008,moritz_2013}, the Hubbard model \cite{moeckel_2008,eckstein_2009}, or sophisticated Holstein models \cite{sentef_2013} were studied so far. Treating an actual femtosecond (fs) laser excitation employed in experimental realizations with optical inter- and / or intraband transitions is challenging \cite{Rameau2016,ligges_2018}. An experimental realization of such models is possible in ultracold atomic gases \cite{joerdens_08, esslinger_10}. For solid materials they represent only a part of the full problem because delocalized Bloch electrons are not included in the model though they are essential in real materials. Therefore, it is important to treat both, the correlated electron states as well as weakly correlated Bloch bands including the interaction between these two electron systems.

In this Letter we analyze such interaction of correlated electrons with Bloch electrons. We investigate hot electron relaxation in the charge density wave / Mott material 1\emph{T}-TaS$_2$by fs time-resolved photoelectron spectroscopy measurements and Boltzmann equation calculations, which treat the interaction of delocalized, propagating electrons in a Bloch band with the correlated electron system. We explain the up to five times longer hot electron lifetimes observed in the correlated, low temperature state as compared to the metallic, high temperature state by a doublon bottleneck in the correlated state. Absence of electronic states in the Mott gap up to an excitation energy set by the Coulomb repulsion $U$ reduces the phase space for electron-electron scattering considerably and increases the hot electron lifetime in the Bloch band. Thus, the observed electron lifetime provides insight into the excited electronic structure in the correlated state, in particular into the interaction strength of Bloch electrons with the localized, strongly correlated electrons.

We perform time-resolved photoelectron spectroscopy on $1T$-TaS$_2$ as described earlier \cite{ligges_2018}. The pump excitation is induced by laser pulses at 1.53~eV photon energy and 50~fs pulse duration, which are generated by regenerative chirped pulse amplification at 250 kHz repetition rate in a commercial Ti:sapphire amplifier (Coherent RegA 9040). Photoelectrons are created by  probe pulses at 6.1~eV photon energy and 100~fs pulse duration, which are obtained by frequency quadrupling in $\beta$-Barium borate crystals and analyzed in normal emission geometry by an electron time-of-flight spectrometer with $\pm0.02$\AA$^{-1}$ parallel momentum resolution and 50~meV effective energy resolution. The cross correlation width of pump and probe pulses is measured on the sample surface at maximum electron kinetic energy and is $110\pm10$~fs. Single crystals $1T$-TaS$_2$ grown as described in detail in \cite{ligges_2018} are cleaved in ultrahigh vacuum at a base pressure of $1\cdot10^{-10}$~mbar.

% figure 1
\begin{figure}
    \centering
        \includegraphics[width=0.85\columnwidth]{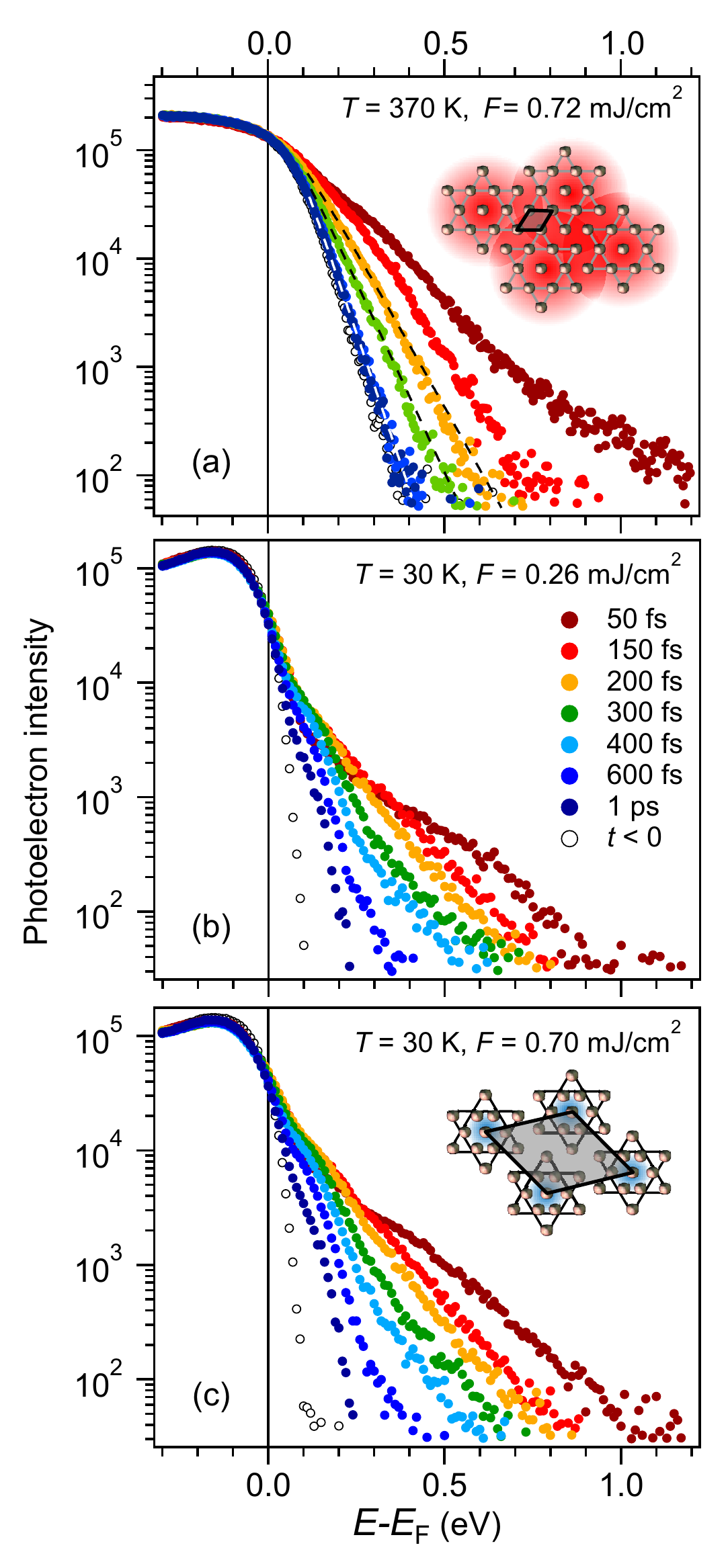}
\caption{Time-resolved photoemission spectra detected in normal emission geometry for the metallic (a) and charge density wave / Mott state (b, c) at the indicated time delay,
temperature, and incident pump fluence $F$. Insets depict the in-plane structure of Ta atoms in the metallic and CDW state. Dashed lines in panel (a) indicate thermalized electron distribution functions at electron temperatures 873~K, 715~K, and 524~ K. All spectra are referenced to $E_{\mathrm{F}}$ of the metallic state.}
    \label{fig:1}
\end{figure}

The investigated material $1T$-TaS$_2$ is a layered transition metal dichalcogenide which presents CDW order of increasing degree of commensurability the lower the temperature $T$
becomes \cite{sipos08}. At $T=370$~K the CDW is incommensurate and a Fermi-Dirac distribution function at the Fermi energy $E_{\mathrm{F}}$ is observed in photoemission spectroscopy, see Fig.~\ref{fig:1}(a), indicating a metallic state. Below $T=180$~K, after having passed through a nearly commensurate (metallic) CDW state, the commensurate CDW forms and the Ta atoms rearrange into clusters out of 13 Ta atoms each in a $\sqrt{13}\times\sqrt{13}$ reconstructed, triangular lattice, as depicted in Fig.~\ref{fig:1}(c). The transition into the commensurate CDW is accompanied by an order of magnitude increase in the electrical resistivity and a loss of spectral weight at $E_{\mathrm{F}}$, see Fig.~\ref{fig:1}(b,c) and \cite{perfetti06}, indicating a Mott transition of the half-filled CDW subband close to $E_{\mathrm{F}}$ \cite{fazekas79,sipos08}. This concept has been challenged recently by the proposal of orbital ordering leading to the insulating state \cite{Ritschel15}. Furthermore, variations in stacking are reported to couple electronic states of adjacent layers \cite{Ngankeu17,lee19}. The broken symmetry ground states of 1\emph{T}-TaS$_2$ provide rich time-dependent structural and electronic dynamics which have led to various ultrafast
experiments which provide manyfold new insights into the complex interplay of lattice and electrons in correlated systems \cite{demsar02,perfetti06,Eichberger2010,hellmann_10, petersen_11,stojchevska14,Vogelgesang2017,ligges_2018}. Here, we exploit the opportunity to compare the ultrafast electron dynamics for a metallic state with the CDW / Mott state in a single material. As such 1\emph{T}-TaS$_2$ serves as a model system to showcase the doublon bottleneck effect, as detailed below.

Fig.~\ref{fig:1} presents photoelectron spectra on a logarithmic intensity axis for the incommensurate, metallic CDW state at $T=370$~K (a) and the commensurate, insulating CDW state
at $T=30$~K (b,c) for different pump-probe time delays $t$. In the metallic case the spectra follow a thermalized Fermi-Dirac distribution function
at $t\geq200$~fs as indicated by the dashed lines in panel (a). With increasing $t$ the respective electron temperature decreases, which is explained
by energy transfer to phonons \cite{Eichberger2010}. At earlier delays weak deviations from a thermalized distribution are identified. In the insulating state the spectra exhibit a more involved behavior. Overall, the electron distribution relaxes towards lower energy with increasing $t$. However, no simple description using a thermal distribution as for 370~K succeeds. Spectra at different $t$ vary weakly up to $E-E_{\mathrm{F}}=0.1$~eV and fan out toward higher energy. Until $t=400$~fs significant electron population at $E-E_{\mathrm{F}}=0.5$~eV is observed. For later $t$ relaxation towards $E-E_{\mathrm{F}}=0.1$~eV is found. These effects are observed for different incident pump fluence $F$, panel (b,c), and are more intense for higher $F$.

\begin{figure}
    \centering
        \includegraphics[width=0.85\columnwidth]{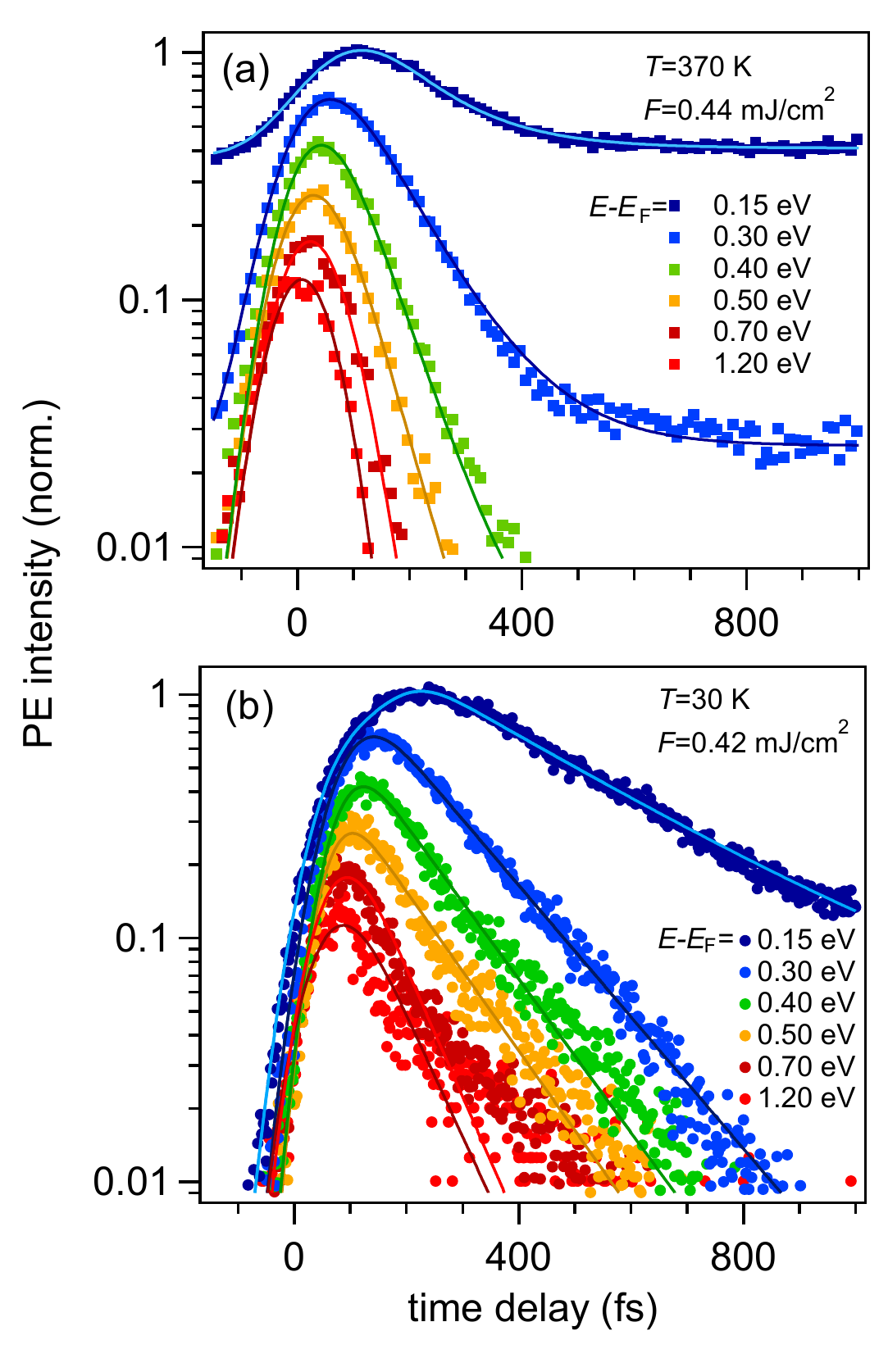}
\caption{Photoelectron intensity as a function of time delay at indicated energies for the
metallic (a) and the CDW / Mott state (b). Solid lines represent fits, see text.}
    \label{fig:2}
\end{figure}

\begin{figure}
    \centering
        \includegraphics[width=0.99\columnwidth]{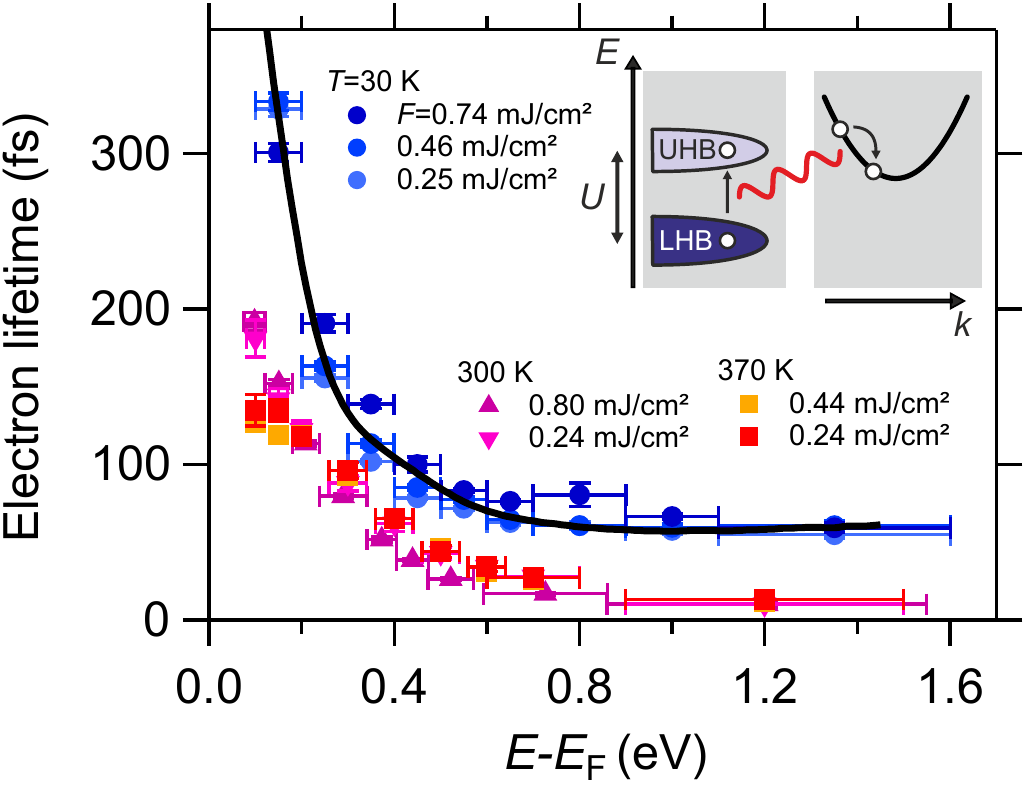}
\caption{Hot electron lifetimes $\tau$ as a function of electron energy determined by fitting
the time-dependent photoelectron intensities for different equilibrium temperatures as indicated.
Horizontal error bars indicate the analyzed, integrated spectral width.
The solid line is the result of the Boltzmann equation calculation, see the supplement.
The inset shows the interaction considered in these calculations:
Relaxation of a delocalized hot electron in a Bloch band ${\cal E}_\mathbf{k}$ mediated by
excitation of a localized doublon-holon pair of energy $\approx U$ in the Fermi-Hubbard system.}
    \label{fig:3}
\end{figure}

For further analysis we turn to the time-dependent photoelectron intensity as a function
of $E-E_{\mathrm{F}}$, because the electron dynamics in the metallic and insulating states
can be analyzed quantitatively in terms of an energy-dependent relaxation time
$\tau(E-E_{\mathrm{F}})$.
Fig.~\ref{fig:2}(a) shows the time-dependent photoelectron intensity at $T=370$~K,
which is %are
characterized by an decreasing $\tau$ with increasing $E-E_{\mathrm{F}}$.
The finite intensity for 0.15 and 0.30~eV at $t>1$~ps originates from thermally populated
states close to $E_{\mathrm{F}}$.
Such qualitative behavior is indeed known for metals because of the increasing phase
space for electron-electron scattering for growing $E-E_{\mathrm{F}}$ \cite{chulkov_2006,Bauer15}.
As depicted in Fig.~\ref{fig:2}(b) $\tau$ decreases with $E-E_{\mathrm{F}}$ also for $T=30$~K,
however much weaker than for the metallic case.
At $E-E_{\mathrm{F}}=1.2$~eV, for example, the relaxation is at 30~K clearly slower than
at 370~K.
We analyze the electron dynamics by fitting the data to
%
% equation 1
\begin{equation}
N(t,E)=\left[\left(N_0 e^{-t/\tau(E)}+N_1\right)\cdot\Theta(t)\right]\otimes g(t)+N_2.
    \label{eq:1}
\end{equation}
Here $\Theta$ is the Heaviside function, $g(t)$ is a Gaussian function representing the cross
correlation width of pump and probe laser pulses, $N_1$ is the population at the asymptotic
value for $t>0$, and $N_2$ is an offset.
The latter two are non-zero for metallic states at 0.15 and 0.3 eV, see Fig.~\ref{fig:2}(a).
The determined lifetimes $\tau(E)$ are depicted in Fig.~\ref{fig:3} for different pump fluences.
While for all data sets lower energy electrons exhibit larger lifetimes than higher energy electrons,
we identify a systematic dependence in $\tau(E)$ on whether the material is in the
metallic state at $T=$300 and 370~K or in the commensurate CDW / Mott state at 30~K.
In the latter case $\tau(E)$ lies well above the values for the metallic state independent
on the chosen $F$. At the highest energy analyzed the determined $\tau$ differ for 30 and 370~K by a factor of five.

% figure 2

% figure 3

In order to illustrate the distinctive qualitative features of the relaxation dynamics,
we consider the following simple model.
The cold electrons in the Mott insulator state are described by the creation and annihilation
operators $\hat c_{\mu,s}^\dagger$ and $\hat c_{\nu,s}$ with spin $s\in\{\uparrow,\downarrow\}$
at the lattice sites $\mu$ and $\nu$.
We model their internal dynamics via the Fermi-Hubbard Hamiltonian \cite{Hubbard63}
\begin{equation}
\label{Fermi-Hubbard}
\hat H_{\rm FH}=-\sum\limits_{\mu,\nu,s}J_{\mu\nu}\hat c_{\mu,s}^\dagger\hat c_{\nu,s}
+U\sum\limits_\mu\hat n_\mu^\uparrow\hat n_\mu^\downarrow
\,,
%\ea
\end{equation}
with the hopping matrix $J_{\mu\nu}$, the on-site repulsion $U$, and the particle numbers
$\hat n_{\mu}^s=\hat c_{\mu,s}^\dagger\hat c_{\mu,s}$.

The hot electrons are described by the operators $\hat a_{\mu,s}^\dagger$ and $\hat a_{\nu,s}$.
In view of the observed fluence independence of their relaxation, see Fig.~\ref{fig:3},
we neglect their interactions among each other and describe their internal dynamics via the
free-electron approximation $\hat H_{\rm free}=
\sum_{\mathbf{k},s}{\cal E}_\mathbf{k} \hat a^\dagger_{\mathbf{k},s}\hat a_{\mathbf{k},s}$
with the Bloch band energies ${\cal E}_\mathbf{k}$.
Since the hot electrons are observed not to populate the upper Hubbard band during their
relaxation \cite{ligges_2018}, we neglect direct transitions between them
and the cold electrons, i.e., we do not include transition terms like
$\hat a_{\mu,s}^\dagger\hat c_{\nu,s}+\rm h.c.$.
Nevertheless, there are interactions (e.g., Coulomb) between them, as described by
\begin{equation}
%\bea
\hat H_{\rm int}=
\sum\limits_{\mu,\nu,s,s'}V_{\mu,\nu}^{s,s'}
\hat c_{\mu,s}^\dagger\hat c_{\mu,s}\hat a_{\nu,s'}^\dagger\hat a_{\nu,s'}
%\hat n_{\mu,s}\hat N_{\nu,s'}
\,,
%\ea
\end{equation}
with the interaction matrix elements $V_{\mu,\nu}^{s,s'}$.

As the total Hamiltonian $\hat H=\hat H_{\rm FH}+\hat H_{\rm free}+\hat H_{\rm int}$ cannot be solved exactly, we have to employ suitable approximations.
For weak interactions $U$ and $V_{\mu,\nu}^{s,s'}$, one may employ standard perturbation theory.
%
%The weak interactions $V_{\mu,\nu}^{s,s'}$ between the hot and cold electrons are treated
%within second-order perturbation theory.
%
However, this is not possible in the strongly correlated Mott state, so we use the method of the hierarchy of correlations instead, see, e.g.,
\cite{navez_10,QNS12,QKNS14,KNQS14,NQS14,NQS16}. To this end, we start with a mean-field approximation of the Mott state $\hat\varrho_\mu\approx(\ket{\uparrow}\bra{\uparrow}+\ket{\downarrow}\bra{\downarrow})/2$ without spin ordering, because we have a triangular lattice which is not bi-partite and
thus prevents anti-ferromagnetic ordering due to spin frustration. Note that we consider the propagation and interaction of the hot and cold electrons within a single layer, i.e. we treat them within a 2D model.

Following a strategy analogous to Ref.~\cite{queisser_18}, we may derive the Boltzmann equation describing the evolution of the distribution functions
$f_{\mathbf{k},s}=\langle\hat a_{\mathbf{k},s}^\dagger\hat a_{\mathbf{k},s}\rangle$ of the hot electrons due to their interaction with the cold electrons in the Mott insulator state.
\begin{eqnarray}
%\bea
\partial_t f_{\mathbf{k},s}=-%\frac{\pi}{2\hbar}
\sum_{s'}\int_{\mathbf{p}}\int_{\mathbf{q}}\;
\left|V_\mathbf{q}^{s,s'}\right|^2
M_{\mathbf{p},\mathbf{q}}
%\left(\frac{J_\mathbf{p+q}-J_\mathbf{p}}{U}\right)^2
\times
\nn
\delta\left({\cal E}_\mathbf{k}-{\cal E}_\mathbf{k-q}-E_\mathbf{p+q}^++E_\mathbf{p}^-\right)
\times
\nonumber\\
\left[
f_{\mathbf{k},s}
\left(1-f_{\mathbf{k-q},s}\right)
\left(1-f^+_{\mathbf{p+q},s'}\right),
f^-_{\mathbf{p},s'}
-{\rm inverse}
\right]
\,.
%\ea
\label{Boltzmann}
\end{eqnarray}
The integrals over momenta $\mathbf p$ and $\mathbf q$ cover the entire Brillouin zone.
As usual, $V_\mathbf{q}^{s,s'}$ is the Fourier transform of the Coulomb interaction matrix
$V_{\mu,\nu}^{s,s'}$ evaluated at the momentum transfer $\mathbf q$.
The above channel describes the inelastic scattering of a hot electron from initial
$\mathbf{k}$ to final momentum $\mathbf{k}-\mathbf{q}$ while creating a doublon-holon
pair with momenta $\mathbf{p}+\mathbf{q}$ and $\mathbf{p}$,
described by their distribution functions $f^+_{\mathbf{p+q},s'}$ and $f^-_{\mathbf{p},s'}$
as well as the matrix elements $M_{\mathbf{p},\mathbf{q}}$, plus the inverse process.
The delta function in the second line of~\eqref{Boltzmann} corresponds
to energy conservation with the doublon and holon excitation energies
\begin{equation}
%\bea
\label{quasi-particle-energies}
E_\mathbf{p}^\pm=\frac12\left(J_\mathbf{p}+U\pm\sqrt{J_\mathbf{p}^2+U^2}\right)
\,,
%\ea
\end{equation}
where $J_\mathbf{p}$ is the Fourier transform of the matrix $J_{\mu\nu}$.

Motivated by the experimentally observed, ultrashort doublon lifetimes \cite{ligges_2018},
we neglect pre-existing doublon-holon excitations in the Mott insulator state,
such that the above relaxation channel~\eqref{Boltzmann} is dominant.
Thus, we may employ the standard relaxation time approximation
$f_{\mathbf{k-q},s}\approx0$,
$f^+_{\mathbf{p+q},s'}\approx0$, and
$f^-_{\mathbf{p},s'}\approx1$ which gives
$\partial_t f_{\mathbf{k},s}=-f_{\mathbf{k},s}/\tau_{\mathbf{k}}$.
As further approximations, we describe the energies of the hot electrons by a parabolic
dispersion ${\cal E}_\mathbf{k}\approx{\cal E}_0+\mathbf{k}^2/(2m_*)$ with the effective mass
$m_*$ and assume that the interaction matrix $V_{\mu,\nu}^{s,s'}$ is dominated by the local
(on-site) term $V_{\mu,\nu}^{s,s'}\approx V\delta_{\mu\nu}$ in analogy to~\eqref{Fermi-Hubbard}.
Under these assumptions we may calculate the relaxation time $\tau_{\mathbf{k}}$.
The qualitative $\mathbf k$-dependence of $\tau_{\mathbf{k}}$ can be understood in terms
of phase space arguments in analogy to the metallic state.
As a peculiarity of two spatial dimensions, the energy ${\cal E}_\mathbf{k-q}$
and the ``volume'' factor of the $\mathbf q$-integral in~\eqref{Boltzmann}
are both quadratic in $\mathbf q$.
Thus, they effectively cancel each other and $\tau_{\mathbf{k}}$ becomes approximately
independent of $\mathbf k$ for large energies, i.e., large $\mathbf k$.

For the strongly interacting limit $U\gg J$ (i.e., deep within the Mott insulating phase),
we may further approximate $E_\mathbf{p+q}^+-E_\mathbf{p}^-\approx U$.
Then we find that the plateau of the relaxation rate $\tau_\mathbf{k}$
at high energies roughly scales with $V^2m_*\ell^2J^2/U^2$,
%
%\begin{equation}
%\bea
%\label{eq:decayrate}
%\frac{\hbar}{\tau_\mathbf{k}}=\ord\left(m_*\ell^2J^2/\hbar^2\right),
%\ea
%\end{equation}
%
where $\ell$ is the lattice spacing
(i.e., the distance between neighboring ``David stars'' in Fig.~\ref{fig:1}c).
%, see Fig.~\ref{fig:3}, inset for illustration.
For smaller energies, however, the available phase space shrinks and eventually vanishes:
If the initial energy ${\cal E}_\mathbf{k}$ of the hot electron becomes too small to
create a doublon-holon pair, i.e., $\mathbf{k}^2/(2m_*)<U$, the channel~\eqref{Boltzmann}
closes and $1/\tau_{\mathbf{k}}$ vanishes.
These phase space arguments yield an approximate step-function behavior of $1/\tau_{\mathbf{k}}$.
Decay of a hot electron in the Bloch band is mediated by excitation of a doublon-holon pair and this
coupling imposes a bottleneck for the relaxation, see Fig.~\ref{fig:3}.
There will be further relaxation channels due to coupling to phonons which will result in a
non-zero relaxation rate $1/\tau_{\mathbf{k}}$ at energies below $U$ leading to lattice heating
as observed experimentally \cite{Eichberger2010}.

%Following Eq.~\ref{eq:decayrate}, we estimate $J(m_*)\approx\sqrt{1.87/m_*}$ for $\ell=3.36$\AA \cite{bovet_2003} and $\tau =60$~fs at 1.2~eV, see Fig.~\ref{fig:3}. In the CDW state back folding of the Bloch band \cite{yu_PRB17} increases $m_*$ compared to the metallic state. Assuming, e.g., $m_*=10m_e$, $m_e$ is the free electron mass at rest, we obtain $J=0.43$~eV supporting the $U/J\sim1$ limit considered for the material under study \cite{ligges_2018}. Due to the square root dependence in $J(m_*)$ this limit is maintained for a wide range of $m_*$. Our model description also provides a fit of $\tau(\cal E)$ as depicted in Fig.~\ref{fig:3}. As detailed in the supplemental material this fit provides values for ${\cal E}_0$ and the interaction $V$ between the Bloch band and the Hubbard electrons. For $m_*=10m_e$ assumed above we find...

%The presence of the doublon bottleneck explains the previously observed absence of refilling of the doublon state by secondary electrons \cite{ligges_2018}. Since the relaxation time of hot electrons $\tau({\cal E)}\geq60$~fs is considerably slower than the decay of the doublon with $\tau=\hbar/J\sim 2$~fs, a doublon populated by a secondary electron will decay with this ultrafast decay time, which suppresses its observation under the given experimental conditions.

In order to arrive at a more quantitative comparison, we have to specify the relevant model parameters such as $J$ and $U$ etc.
Unfortunately, their precise value is still a somewhat open question, so we assume potentially realistic values of $U=0.35~\rm eV$ and $J=0.05~\rm eV$ as a working hypothesis,
cf.~Ref.~\cite{ligges_2018}.
The Bloch band ${\cal E}_\mathbf{k}\approx{\cal E}_0+\mathbf{k}^2/(2m_*)$ is described by the two parameters ${\cal E}_0$ and $m_*$.
Since photoelectrons are observed down to quite low energies, we assume ${\cal E}_0=0$.
Other values of ${\cal E}_0$ would just shift the curve $\tau(E)$ horizontally and thus do not affect the height of the plateau at high energies.
However, the effective mass $m_*$ does also affect the functional form of $\tau(E)$ as well as the plateau height, remember the rough scaling law
$1/\tau_\mathbf{k}\sim V^2m_*\ell^2J^2/U^2$ described above.
In view of the delocalized and quasi-free nature of the hot electrons, we assume $m_*=m_e$.
Then, using the approximation $V_{\mu,\nu}^{s,s'}\approx V\delta_{\mu\nu}$ mentioned above, we may obtain the remaining unknown parameter $V$
by fitting the energy dependent relaxation rate $\tau(E)$, especially the plateau height, see Fig.~\ref{fig:3}, which gives
$V\approx 0.085~\rm eV$.

%\bigskip

%Then, using the approximations
%${\cal E}_\mathbf{k}\approx{\cal E}_0+\mathbf{k}^2/(2m_*)$
%and
%$V_{\mu,\nu}^{s,s'}\approx V\delta_{\mu\nu}$
%mentioned above, we may obtain the remaining unknown parameters
%${\cal E}_0$, $m_*$ and $V$ by fitting the energy dependent relaxation rate $\tau(E)$,
%see Fig.~\ref{fig:3}.
%
%Varying the Bloch band minimum ${\cal E}_0$ just moves the curve $\tau(E)$
%to the left or right while $V^2$ acts as a global factor.
%The effective mass $m_*$ does also affect the functional form of $\tau(E)$
%such as its curvature.
%
%However, other relaxation channels which are not included here
%(such as the coupling to phonons, which is more pronounced at lower energies)
%may also change this curvature -- such that the most robust value is given by the
%height of the plateau at large energies, which allows us to determine $V$.
%
%Following this strategy, the best fit is obtained for
%${\cal E}_0=-0.18~\rm eV$,
%$m_*=0.15~m_e$
%and
%$V=0.22~U$,
%see Fig.~\ref{fig:3}.

A value of $V$ below $U$ is quite natural as $U$ is determined by the Coulomb overlap integral between the same charge density distributions (up to the opposite spin)
of the cold electrons in the Mott state while $V$ corresponds to the overlap between the charge density distributions of the cold and the hot (Bloch band) electrons.
At small energies, the fit in Fig.~\ref{fig:3} should be taken {\em cum grano salis} because there low-energy relaxation channels such as coupling to phonons
(which we have not included here) may become important.
Nevertheless, the result $V<U$ obtained from the plateau at high energies is quite robust, unless extremely small values of $m_*$ are assumed.

%As explained above, the values for ${\cal E}_0$ and $m_*$ should be taken {\em cum grano salis}
%as they can depend on the low-energy relaxation channels such as coupling to phonons
%which we have not included here.
%
%For example, inserting half the effective mass $m_*\to m_*/2$, the fit to the experimental
%data is also not too bad and yields a slightly larger value for $V=0.35~U$,
%consistent with the rough scaling law $1/\tau_\mathbf{k}\sim V^2m_*\ell^2J^2/U^2$ described above.
%
%Nevertheless, the result $V<U$ obtained from the plateau at high energies is quite robust,
%unless extremely small values of $m_*$ are assumed, for which it becomes really hard to
%explain the experimental data realistically.

In conclusion, we have observed hot electron decay in a strongly correlated electron material to exhibit at high energy ${\cal E}>U$ a rather long lifetime, which we assign to its coupling to a secondary doublon-holon excitation acting as a bottleneck in the decay. A model description which considers coupling of delocalized, hot Bloch electrons with the
doublon-holon excitation provides a description of the excited electronic structure, which we expect to have considerable impact in the field of strongly correlated electron materials
in general.

\begin{acknowledgments}
This work was funded by the Deutsche Forschungsgemeinschaft (DFG),
grant \# 278162697 (SFB 1242).
%
%This work was funded by the Deutsche Forschungsgemeinschaft (DFG, German Research Foundation)
%Projektnummer  278162697 - SFB 1242.
\end{acknowledgments}

\providecommand{\noopsort}[1]{}\providecommand{\singleletter}[1]{#1}%

\end{document}